\documentclass{nature}
\usepackage{graphicx}

\usepackage{amsmath}
\usepackage{amssymb}
\begin{document}
\title{Electrostatic quantum dots in silicene }

\author{B. Szafran \footnote{Correspondence to bszafran@agh.edu.pl}, D. \.Zebrowski,  Alina Mre\'{n}ca-Kolasi\'{n}ska, \\
   AGH University of Science and Technology, \\Faculty of Physics and
Applied Computer Science,\\
 al. Mickiewicza 30, 30-059 Krak\'ow, Poland
}

\maketitle

\begin{abstract}
We study electrostatic quantum dot confinement for charge carriers in silicene.
The confinement is formed by vertical electric field surrounding the quantum dot area.
The resulting energy gap in the outside of the quantum dot traps the carriers within,
and the difference of electrostatic potentials on the buckled silicene sublattices produces nonzero carrier masses 
outside the quantum dot. 
We study the electrostatic confinement defined inside a silicene flake with both the atomistic tight-binding approach as well as with the continuum approximation for a circularly symmetric electrostatic potential.
We find localization of the states within the quantum dot  and their decoupling from the edge that makes
the spectrum of the localized states independent of the crystal termination. For an armchair edge of the flake removal
of the intervalley scattering by the electrostatic confinement is found. 
\end{abstract}


\section*{Introduction}

Silicene \cite{silitmdc,chow} is a material similar in crystal and electron structure to graphene \cite{Neto09}
but with enhanced spin-orbit coupling \cite{Liu11,Liu,Ezawa} that makes this two-dimensional medium attractive for
studies of  anomalous-  \cite{Ezawa}, spin- \cite{Liu11} and valley- quantum  Hall effects \cite{Pan14},
giant magnetoresistance \cite{Xu12,Rachel14} and construction of spin-active devices \cite{Zutic04,miso15}.
The crystal structure of a free-standing   silicene is buckled \cite{bu2} with a relative shift of the triangular A and B sublattices in the vertical direction.
The shift allows one to induce and control the energy gap near the charge neutrality point  \cite{ni,Drummond12}.
The silicene was first successfully formed on metallic substrates \cite{Vogt12,Aufray10,Feng12,siedem,osiem,dziewiec}.
For the studies of electron properties of systems based on silicene non-metallic substrates \cite{springer} are needed.
Theoretical studies have been performed for the silicene on insulating AlN \cite{aln}, and semiconducting  transition metal dichalcogenides (TMDCs) \cite{silitmdc,tmdc1,tmdc2}.
An operating room-temperature field effect transistor was recently realized \cite{Tao15} with silicene layer on Al$_2$O$_3$ insulator.  Al$_2$O$_3$  only weakly perturbs the band structure of free-standing silicene near the Dirac points \cite{al2o3}.

In this paper we study formation of an electrostatic quantum dot within the silicene.
 The electrostatic quantum dots \cite{hanson09} allow for precise studies of the carrier-carrier and spin-orbit interaction. 
In graphene the electrostatic confinement is excluded  since the carriers behave like massless Dirac fermions that evade
electrostatic confinement  due to the lack of the energy gap in the dispersion
relation and chiral Klein tunneling that prevents backscattering \cite{k2}.
A local electrostatic potential in graphene can only support quasibound states \cite{qb1,qb2}
of a finite lifetime and cannot permanently trap the charge carriers.
Carrier confinement and storage  can be realized by finite flakes of graphene 
 \cite{gru,zarenia,flake2,flake5,Ezawaf}. However, 
the electron structure of states confined within the flakes depends
strongly on the edge \cite{gru,zarenia} that is hard to control at the formation stage
and cannot be changed once the structure is grown. The electrostatic confinement \cite{hanson09} is free from these limitations. Finite flakes
of silicene as quantum dots were also discussed \cite{kiku,romera,abdelsalam}. 
For the graphene, the energy gap \cite{nrbr} due to the lateral confinement or mass modulation by eg. a substrate allows for formation of quantum dots by external potentials \cite{silvestrov07,han,xliu,nori}. Confinement by inhomogeneity of the magnetic field has also been proposed for graphene \cite{m1,m3}
which removes the edge effects.

The  electrostatic quantum dots studied below are formed by an inhomogeneous vertical electric field. 
We consider a system in which the confinement of the carriers is induced within a region surrounded by strong vertical electric fields [see Fig. 1].
The inhomogeneity of the electric field is translated into position-dependence of the energy gap. Localized states are
 formed within a region of a small energy gap surrounded by  medium of a larger  gap. 
A similar  confinement mechanism has previously been demonstrated for bilayer graphene \cite{pereira},
which also reacts to the vertical electric field by opening the energy gap.
The vertical electric field produces  potential variation at the A and the B sublattices of the buckled silicene [Fig. 1(b)].
In this way the system mimics the idea for potential confinement of neutrinos 
introduced by  Berry and Mondragon \cite{bm}. A potential of a different sign for the components of spinor wave function
was applied \cite{bm} 
that produces a so-called infinite-mass boundary in the limit case of a large potential. 
The infinite-mass boundary condition is applied for phenomenological modeling of graphene flakes with the Dirac equation \cite{bm,zarenia,gru,thomsen,nori}.
The proposed device is a physical realization of this type of the boundary condition. 
Note, that for monolayer TMDCs \cite{tmdc}, materials with hexagonal crystal lattice, the valley degree of freedom 
and strong spin-orbit coupling, formation of electrostatic quantum dots \cite{burkard} is straightforward due to the wide energy gap of the system.
However, these systems are far from the Dirac physics for massless or light carriers.

\section{Theory}

\begin{figure}
 \includegraphics[width=0.8\columnwidth]{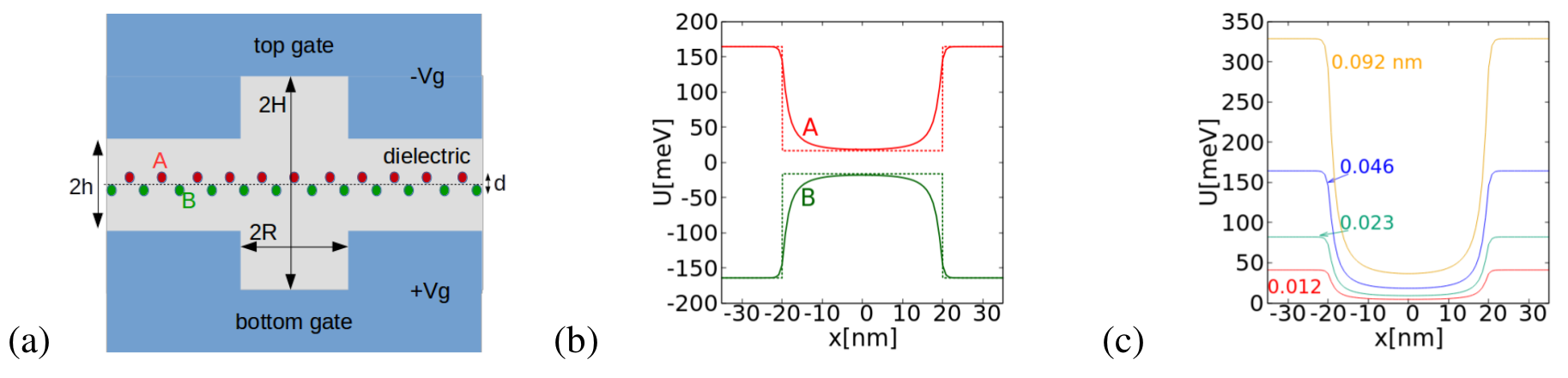}  
\caption{(a) Schematics of the quantum dot device. The silicene layer is embedded within a dielectric 
inside a symmetric double gate system. The distance between the A and B sublattice planes is $d=0.046$ nm. 
The spacing between the gates within the central circular
region of diameter $2R=40$ nm is $2H=28$ nm, and $2h=2.8$ nm outside. (b) The solid lines
show the electrostatic potential for $V_g=\pm 10$ V applied to the gates at
the A and B sublattices as calculated from the Laplace equation.
The dashed lines indicate a rectangular quantum well approximation used in the calculations (see text). 
The cross section of the electrostatic potential in (b) is taken at $y=0$ and $z=\pm d/2$. 
(c) The potential on the A sublattice for the parameters of (b) for varied buckling, i.e. the vertical offset between the A and B sublattices
with the values of $d$ given in the plot in nanometes. The calculations in this work are performed for $d=0.046$ nm. 
  }
\label{wincy}
\end{figure}

\subsection{Model system}
We consider  silicene embedded  
in a center of a dielectric layer sandwiched symmetrically between metal gates [Fig. 1(a)]. The distance between the gates
is $2h=2.8$ nm outside  a circular region of radius $2R=40$ nm, where the spacing 
is increased to $2H=28$ nm. The model device is a symmetric version of an early electrostatic GaAs quantum dot device \cite{ashoori}.  
The electrostatic potential  near the charge neutrality point can be estimated by solution of the Laplace equation
with the Dirichlet boundary conditions at the gates. The solution on the A and B sublattices is shown
in Fig. 1(b) for the gate potential $V_g=10$ V. A potential difference between the sublattices presented in Fig. 1(b)
appears as a result of the buckled crystal structure with the vertical distance $d=0.046$ nm; between the sublattices [see Fig. 1(a)]. The difference  is large outside the central circular area.
Beyond  this area the potential is $U_A= e V_g d/2h$ for the A sublattice and $U_B= - e V_g d/2h$  for the B sublattice [Fig. 1(b)].
Near the center of the circular area the potential is $U_{A}= e V_g d/2H$, $U_B=-U_A$ with the gate potential lever arm increased
by the larger spacing between the gates. 
The bottom of the electrostatic potential in the center of the dot in Fig. 1(b) is flat.
The electrostatic potential  can be approximated by a formula $V_{exact}=eV_g\frac{d}{2h}[1-\exp(-(r/R)^m)]+eV_g\frac{d}{2H}$, 
with $m\in(6,8)$ i.e. 
in the Taylor expansion of the potential the parabolic term corresponding to the harmonic oscillator potential is missing. 
 Therefore, 
in the calculations below we consider a rectangular potential well
model
 \begin{equation}  U_A(r)= \left\{ \begin{matrix}  e V_g \frac{d}{2h} & \; \mathrm{for} \; r>R \\  \ e V_g \frac{d}{2H} & \; \mathrm{for} \; r \leq R \\ \end{matrix} \right.,\end{equation} and $U_B=-U_A$, where $r$ is the in-plane distance from the center of the system. The model potential is plotted with the dotted lines in Fig. 1(b). The results
for the exact electrostatic potential are also discussed below. For the discussion of the confinement potential profile depending on the geometry of the gates see Ref. \cite{szafranbednarek}. Note, that the gate voltage to confinement potential conversion factor depends not only on the
spacing between the gates but on the buckling distance \cite{silitmdc} which varies for different substrates. The confinement potential ar the $A$ sublattices is plotted for varied
values of the buckling distance $d$ given in nanometers. 

\subsection{Atomistic Hamiltonian}
For the atomistic tight-binding modeling we apply the basis \cite{Liu} of $p_z$ orbitals, for which the Hamiltonian reads \cite{Liu,Ezawa,Ezawa12a}
\begin{eqnarray}
H&=&-t\sum_{\langle k,l\rangle \alpha } e^{i\frac{e}{\hbar}\int_{\vec{r_k}}^{\vec{r_l}}\vec {A}\cdot \vec {dl}}  c_{k\alpha}^\dagger c_{l\alpha}  +i\frac{\lambda_{SO}}{3\sqrt{3}}\sum_{\langle \langle k,l\rangle \rangle \alpha, \beta }  e^{i\frac{e}{\hbar}\int_{\vec{r_k}}^{\vec{r_l}}\vec {A}\cdot \vec {dl}} \nu_{kl} c^\dagger_{k\alpha} \sigma^{z}_{\alpha\beta}c_{l\beta} \nonumber \\
\\ && + \sum_{k,\alpha} U({\bf r}_k) c^\dagger_{k\alpha}c_{k\alpha} +\frac{g\mu_B B}{2}\sum_{k,\alpha}  c_{k\alpha}^\dagger \sigma^z_{\alpha,\alpha}  c_{l\alpha}, \nonumber  \label{hb0}
\end{eqnarray}
where 
 $\sigma_z$ is the Pauli spin matrix, 
$c_{k\alpha}^\dagger$ is the electron creation operator at ion $k$ with spin $\alpha$,
the symbols $\langle k,l\rangle $ and $\langle\langle k,l\rangle\rangle $  stand for the pairs of nearest neighbors  and  next nearest neighbors, respectively. 
The first term of the Hamiltonian accounts for the nearest neighbor hopping with $t=1.6$ eV \cite{Liu,Ezawa}. 
 The second term describes
the intrinsic spin-orbit interaction \cite{km} with the sign parameter $\nu_{kl}=+1$  ($\nu_{kl}=-1$) for the counterclockwise (clockwise) next-nearest neighbor hopping and $\lambda_{SO}=3.9$ meV \cite{Liu,Ezawa}.
 The exponents in the first and second sum  introduce the Peierls phase, with the vector potential $\vec{A}$.  The term with $U$ introduces the model electrostatic potential given by Eq. (1). The last term is the spin Zeeman interaction for perpendicular magnetic field, where $\mu_B$
is the Bohr magneton and the electron spin factor is $g=2$.  The applied Hamiltonian is spin diagonal in the basis of $\sigma_z$ eigenstates. 
 We consider the states confined 
within the confinement potential that is defined inside a finite silicene flake containing up to about 72.5 thousands ions. 

\subsection{Continuum approximation}

The continuum approximation to the atomistic Hamiltonian provides the information on the valley index and angular momentum of the confined states.
The continuum Hamiltonian (\ref{hb0}) near the $K$ and $K'$ valleys written 
for the spinor functions $\Psi=(\Psi_A,\Psi_B)^T$
is \cite{Ezawa12a}
\begin{equation} H_\eta = \hbar V_F \left(k_x \tau_x -\eta k_y \tau_y \right)+U(r)\tau_z+\frac{g\mu_B B}{2}\sigma_z -\eta\sigma_z \tau_z \lambda_{SO},
\end{equation}
where the valley index is $\eta=1$ for the $K$ valley and $\eta=-1$  for the $K'$ valley, $\tau_x$, $\tau_y$ and $\tau_z$ are the Pauli
matrices in the sublattice space, ${\bf k}=-i\nabla+\frac{e}{\hbar} \vec{A}$, and $V_F=\frac{{3at}}{2\hbar}$ is the Fermi velocity
with the nearest neighbor distance $a=0.225$ nm. 

For the isotropic potential $U(r)$ and the symmetric gauge $\vec{A}=(-By/2,Bx/2,0)$ the $H_\eta$ Hamiltonian commutes with 
the total angular momentum operator of the form $J_z=L_z {\bf I} +\eta \frac{\hbar}{2} \tau_z$, where $L_z=-i\hbar \frac{\partial }{\partial \phi}$ is the orbital angular momentum operator,
and ${\bf I}$ is the identity matrix. The components of the common $H_\eta$ and $J_z$ eigenstates 
can be put in a separable form \begin{equation} \Psi_\eta=\left (\begin{matrix}  f_A (r) \exp(im\phi) \\ f_B (r) \exp(i(m+\eta)\phi)\end{matrix} \right) \label{psi} \end{equation}
where $m$ is an integer. For the $K'$ valley states we will denote the quantum number by $m'$. The asymptotic behavior of the radial functions for a given $m$ and $\eta$ in the center of the potential is $f_A\sim r^{|m|}$ and $f_B\sim r^{|m+\eta|}$ \cite{thomsen}.
The radial components fulfill the system of equations 
\begin{eqnarray}
&& \left(U_A(r)+\frac{g\mu_B B \sigma_z}{2}\right)f_A\nonumber \\ &+& V_F\left[-\eta\frac{i\hbar}{r}(m+\eta)f_B-i\hbar f'_B-\eta \frac{iBr}{2}f_B \right]=E f_A, \label{e1}\\
&& \left(U_B(r)+\frac{g\mu_B B \sigma_z}{2}\right)f_B\nonumber \\ &+&  V_F\left[ \eta\frac{i\hbar}{r}m f_A-i\hbar f'_A+\eta \frac{iBr}{2}f_A \right]=E f_B, \label{e2}
\end{eqnarray}
which is solved numerically using a finite difference approach. 
The continuum Hamiltonian eigenstates have a definite $z$ component of the spin, the valley index, and the angular momentum. 
Below we label the Hamiltonian eigenstates of $K$ [$K'$] valley with the $j$ [$j'$] angular momentum quantum number,
with $j=m +1/2$ [$j'=m' -1/2$].

In the continuum approach we look for the states localized within the confinement potential of radius $R$  within a finite circular flake of radius $R'$. 
We are interested in the influence of the type of the flake on the states localized within the electrostatic potential well. 
In the continuum approximation at the edge of the flake we apply two types of boundary conditions: the  infinite-mass boundary condition $\frac{f_B}{f_A}_{|r=R'}=i$ \cite{bm,zarenia,gru}
and the zigzag boundary condition for which one of the components of the wave function vanishes at the end of the flake $r=R'$. 
The zigzag edge supports localized states with zero energy at $V_g=0$.
 With the infinite mass boundary conditions the zero energy states \cite{kiku} are missing and  the low-energy states  are extended over the interior of the flake.
The infinite mass and zigzag boundary conditions preserve the valley index as a quantum number. The maximal mixing of the valleys appears with the armchair
edge of the flake. The latter is considered with the atomistic tight-binding approach.

\section{Results}

\begin{figure}

 \includegraphics[height=0.8\columnwidth]{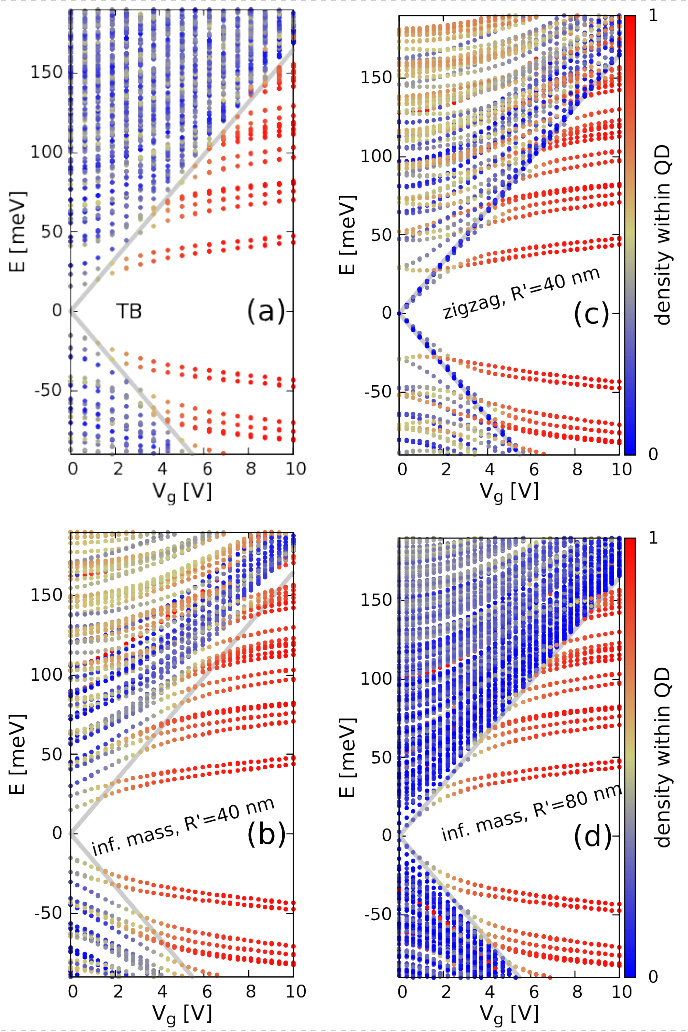} 

\caption{Energy levels of a silicene flake with a circular quantum dot of radius $R=20$ nm defined in its
center by the inhomogeneous vertical electric field as functions of the gate voltage applied as in Fig. 1. 
The plot (a) shows the results of the atomistic tight binding approach for the armchair hexagonal flake of side length 43 nm without the spin-orbit coupling.
Plots (b-d) were obtained with the continuum approach for the confinement potential defined of radius $R=20$ nm within a  circular flake of radius $R'=40$ nm (b,c) and $R'=80$ nm (d). 
The infinite mass boundary conditions were applied at the edge of the flake in (b) and (d) and zigzag boundary conditions in (c).
The color of the lines indicates the part of the probability density that is localized at a distance of 1.1 $R$ from the center of the dot.
The thick gray lines show the electrostatic potential energy at the A (the upper gray line with positive energy) and B (the lower gray line)
outside of the quantum dot.  The results are obtained for perpendicular magnetic field $B=0.5$ T and the spin-orbit coupling is neglected. }
\label{ely}
\end{figure}

\begin{figure}

\includegraphics[width=.7\columnwidth]{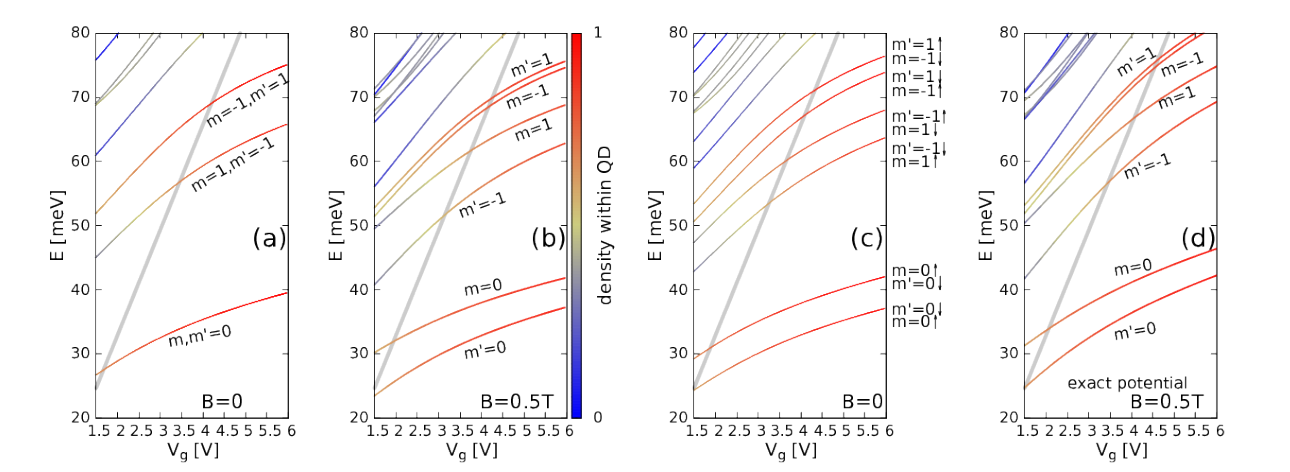}

\caption{The solution of the Dirac equation for the quantum dot of radius $R=20$ within a circular flake with  $R'=40$ nm and  infinite mass boundary conditions at the flakes edge.
The color of the lines 
shows the localization of the electron within $1.1R$ from the center of the system and the scale is given to the right of (b).  
In the figure we mark the angular momentum quantum number for the A component,
$m$ for the energy levels belonging to the $K$ valley and $m'$ for the ones in $K'$ valley. 
 In (a) and (b) the spin-orbit coupling
is absent, the applied magnetic fields  are $B=0$ (a) and $B=0.5$ T (b).  Plot (b) is a zoom of 
parameters of Fig. 2(b).  In (c) $B=0$ and the spin-orbit coupling is switched on. In (c)  $\uparrow$, $\downarrow$ stand for the $z$ component of the spin.
In (a) the energy levels are fourfold degenerate: with respect to the valley and the spin.
In (b) the valley and spin degeneracy is lifted, but on the plot one resolves
only the valley splitting, the Zeeman effect energy is too small to resolve the splitting of the lines.
In (d) we plot the results obtained for the exact electrostatic potential of Fig. 1(b).
Other plots (a-c) are obtained for the rectangular potential well of Eq. (1) as elsewhere in this work. 
 }
 
 \label{zzom}
\end{figure}

\begin{figure}

\includegraphics[width=.6\columnwidth]{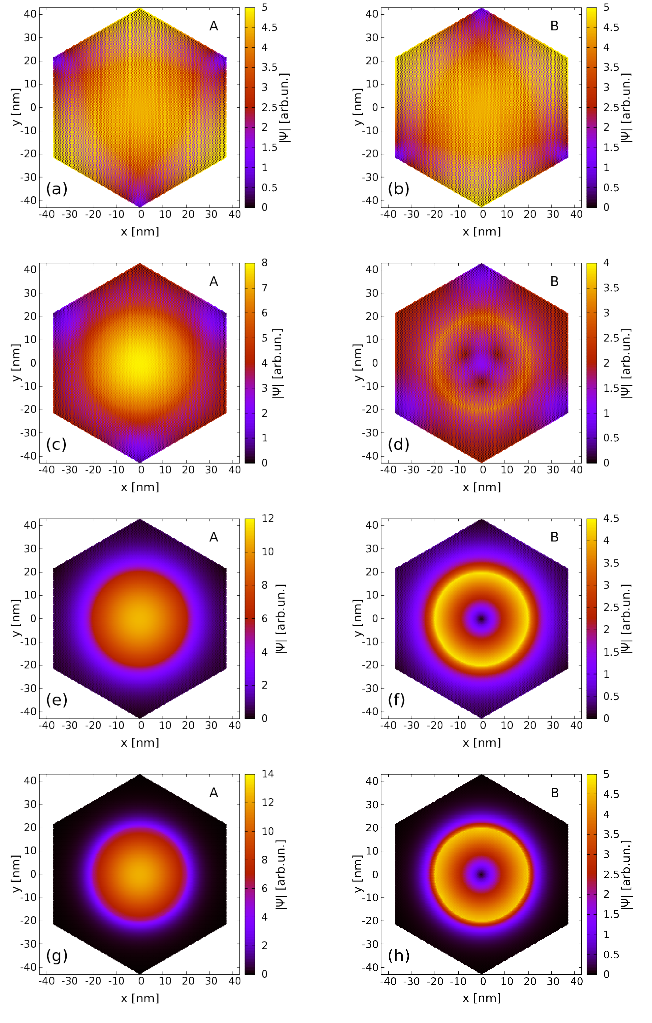}

 \caption{The absolute value of the tight-binding wave function  at A (left column) and B (right column) 
for the lowest conduction band energy level at $B=0.5$ T for $V_g=0$ (a,b), $V_g=1.875$ V (c,d), $V_g=5$ V (e,f) and $V_g=10$ V (g,h).
The results are obtained for a hexagonal armchair flake of side length 43 nm.
In the continuum approach the localized ground-state is a $K'$ valley $j'=-1/2$ state with orbital angular momentum 0 and -1 for the A and B sublattice components, respectively. 
} \label{futb}
\end{figure}

\begin{figure}
\includegraphics[width=.8\columnwidth]{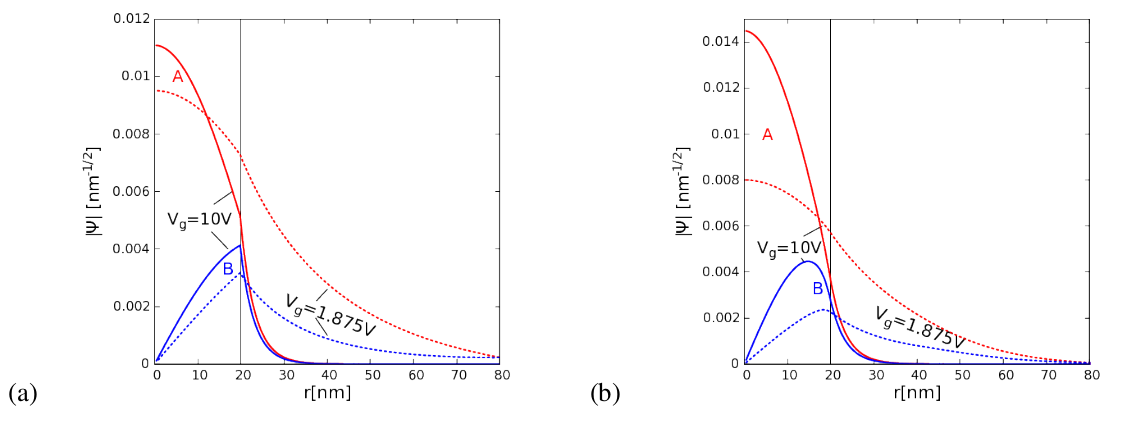}  

\caption
{Absolute value of the radial functions for the $K'$, $j'=-1/2$ state at 0.5T for $V_g=10$ V (solid lines) and $V_g=1.875$ V (dashed lines) for the continuum Hamiltonian.
The A (B) sublattice component is plotted with the red (blue) line. The vertical solid line indicates the radius of the quantum well $R=20$ nm defined within
the flake of 80 nm. 
The applied normalization condition is $\int r\left( |\Psi_A|^2+|\Psi_B|^2\right) dr=1$. The infinite mass boundary conditions are applied at the end of the flake at $r=R'=80$ nm. 
Panel (a) shows the results for the rectangular potential well [Eq.(1)] and (b) for the smooth electrostatic potential of Fig. 1(b).
} \label{radial}
\end{figure}

Figure 2 shows the energy spectrum  and the localization of energy levels obtained with the atomistic tight binding [Fig. 2(a)] and with the continuum approach [Fig. 2(b-d)] 
as functions of the gate voltage. In this plot the spin-orbit interaction was switched off.  A vertical magnetic field of 0.5 T is applied, for which splitting
of energy levels with respect to the valley but not with respect to the spin is visible on the scale of Fig. 2.
 One observes the splitting
of the energy levels with respect to the orbital angular momentum in the external magnetic field. 

In the atomistic tight-binding approach [Fig.2 (a)]  a hexagonal flake of side length 43 nm and an armchair boundary were taken.
For the continuum approach  a circular flake of radius  $R'=2R=40$ nm (b,c) and $R'=4R=80$ nm (d) were studied. In Fig. 2(b,d) the infinite mass boundary 
condition is applied at the end of the flake and in Fig. 2(c) the zigzag boundary condition is assumed.
The energy levels in this area get localized inside the quantum dot -- see the color of the points
that indicate the localization of the electron probability density inside the quantum dot. 
The zigzag edge applied in Fig. 2(c) supports
the edge-localized energy levels which correspond to zero energy in the absence of external fields.
The edge energy levels for the zigzag flake in Fig. 2(c) are split
by the gate potential \cite{kiku,abdelsalam}. The energy of  the edge states \cite{kiku} follow 
 the potential energy at the separate sublattices  outside  the quantum dot.
The edge states are missing for  the armchair edge of the  hexagonal flake adopted for the tight-binding calculations in Fig. 2(a)
and for the infinite-mass boundary condition adopted in the continuum model in Fig. 2(b,d).
 For the larger circular flake  [Fig. 2(d)]  the spacing between the energy levels
localized outside of the dot is decreased, but the same spectrum of the localized states is found.
We can see in all the panels of Fig. 2 that the energy spectra of localized states obtained by the atomistic and continuum approaches with varied boundary conditions  become similar for larger $V_g$. The localized states are found in between the two thick gray lines that show potential energy at the A and B sublattices
outside the quantum dot  $U=\pm e V_g d/2h$. A perfect agreement between the energies of the localized states in the  tight-binding and Dirac models is obtained for the energy levels that are the closest to the charge neutrality point $(E=0)$. For the energy levels that are closer to edge states energy [cf. Fig. 2(a) and Fig. 2(b) for $E>100$ near $V_g=10$ V], the wave functions of the localized states penetrate into the region outside of the quantum dot. The external region 
is different in all the plots of Fig. 2, hence the resolvable difference of the energy levels. 
The spectrum of the zigzag flake [Fig. 2(c)] indicates that the confinement of subsequent states
within the quantum dot area appears with the crossing of the confined energy levels with the edge states \cite{kiku} which
shift linearly with the external potential. The edge states and thus the crossings are missing 
in the results obtained with the armchair edge [Fig. 2(a)], and the infinite-mass boundary conditions [Fig. 2(b,d)].

The effects of the spin-orbit coupling and the results for the exact confinement potential are given in Fig. 3. 
The plot of Fig. 3(b) -- with the external field 0.5 T and without the spin-orbit interaction is the zoom of Fig. 2(b).
The Zeeman spin splitting is still not resolved at this energy scale, but  the splitting of the energy states
with respect to the valley is evident. The K (K') valley states with the indicated angular momentum quantum number $m$ ($m'$) for the A sublattice is given 
in the Figure.  In Fig. 3(b) all the energy levels are nearly degenerate with respect to the spin.
 For comparison the result for 0 T is plotted in Fig. 3(a), where all the energy levels are strictly degenerate. The degeneracy is fourfold:
with respect to both the spin and the valley. The results with the intrinsic spin orbit coupling are displayed in Fig. 3(c) for $B=0$. 
The intrinsic spin-orbit interaction introduces an effective valley-dependent magnetic field which forms spin-valley doublets. The energy effects
of the splitting is comparable to the external magnetic field of 0.5 T given in Fig. 3(b). 

The results of the present manuscript were obtained with the rectangular quantum well potential [Eq. (1)] approximation to the actual electrostatic potential [see Fig. 1(b)].
The results for the rectangular potential [Fig. 3(b)] can be compared to the ones with the exact potential [Fig. 3(d)].
The energy levels for the exact potential are shifted up on the energy scale -- since the rectangular potential well is a lower bound to the exact potential [cf. Fig. 1(b)]. However, the order of the energy levels and the relative spacings obtained with the exact potential are close to the ones obtained for the quantum well ansatz.

Figure~2 demonstrates that the dot-localized states are insensitive to the type of the edge and the size of the flake,
which results from their decoupling from the edge. In particular, the valley mixing by the armchair edge is removed.
The removal of the valley mixing has distinct consequences for the electron wave functions as described within the atomistic approach.
Figure 4 shows the absolute value of the probability amplitude at the A (left column) and B (right column) sublattices for
varied values of the gate voltage and the lowest-energy conduction-band state for $B=0.5$ T. 
For $V_g=0$ the electron density at both the sublattices undergoes rapid oscillations which result from contributions
from both valleys --  distant in the wave vector space -- to the electron wave function in the real space. 
In presence of the valley mixing the low-energy wave function for sublattice A can be written as a superposition 
$\Psi_{A}({\bf r})=\exp(i {\bf K} \cdot {\bf r})\phi_{A}({\bf r})+\exp(i {\bf K'} \cdot {\bf r})\phi_{A'}({\bf r})$.
 The probability density is then $|\Psi_{A}({\bf r})|^2=|\phi_{A}({\bf r})|^2+|\phi_{A'}({\bf r})|^2 +2 \Re \left(\phi_A^*({\bf r})\phi_{A'}({\bf r})\exp(i({\bf K'}-{\bf K})\cdot{\bf r})\right)$. 
Due to the large distance between ${\bf K}$ and ${\bf K}'$ in the reciprocal space the exponent term induces rapid variation of the density from one atom to the other even when $|\phi_A|^2$ and $|\phi_A'|^2$ densities are smooth.
A smooth $|\Psi_A|$ amplitude can only be obtained provided that one of the valley components $\phi_A$ or $\phi_{A'}$ is zero.
Figure 4 shows that indeed as the gate voltage is increased the rapid oscillations of the density disappear. The valley mixing disappears
along with the coupling to the edge.

In Fig. 4 a circular symmetry of the confinement potential is reproduced by the 
electron density for larger $V_g$.   In the lowest-energy conduction band state that we plot in Fig. 4 the density is locally maximal in the center of the quantum dot in the A sublattice.
For the B sublattice a zero of the density is found. In the continuum approach 
the ground state at $B>0$ corresponds to $K'$ valley with the total angular momentum $j'=-1/2$ or $m'=0$ in Eq. (\ref{psi}). The A component
of the wave function corresponds to an $s$ state and the B component to a $p$ state, which agrees with the results of Fig. 4(g,h). 
For $V_g=10$ V  the electron density far from the dot disappears. A penetration of the electron density outside the nominal radius of the dot is still present, but short range.

In Figure 5(a) we plot the absolute value of the wave function for the same state as obtained with the continuum approach 
with the infinite-mass boundary condition at the end of the flake $R'=80$ nm. 
The vanishing derivative of the probability amplitude at $r=0$ is found for the A sublattice and a linear behavior for the radial function on the B sublattice.
Equations (\ref{e1},\ref{e2}) translate the potential step 
into a jump in the derivative of the radial functions. $V_g$ shifts most of the probability amplitude of the lowest-energy conduction band states to the A sublattice
(see also Fig. 4). However, for large $V_g$ the radial functions for both sublattices tend to the same amplitude
(see Fig. 5(a)), since in the limit of infinite $V_g$ the variation of the electrostatic potential at the outside of the dot
induces  an  infinite-mass boundary at $r=R$, which implies equal amplitudes of the wave functions therein \cite{bm,zarenia,gru}.
The results obtained with the exact potential are given in Fig. 5(b). The derivatives of the wave function are continuous
for the smooth potential variation. The maxima of the wave function amplitude on the B sublattice no longer exactly coincide with $R$
for the exact potential. Also,  Fig. 1(b) shows that the energy difference between the sublattices is larger for the exact potential
than in the rectangular quantum well approximation.

\section{Summary and Conclusions}
We found formation of states localized by external electrostatic potential within a silicene flake. The potential
used for this purpose results from the inhomogeneity of the vertical electric field that induces an energy gap outside the quantum dot
and the buckling of the silicene surface. 
The energy spectrum for a finite flake can be separated into quantum-dot localized states and the states delocalized over the rest of the flake. The localized and delocalized states
appear in separate parts of the energy spectrum limited  by the electrostatic potential on the separate sublattices of silicene. 
We have demonstrated that the states localized within the quantum dot are separated from the edge and independent of the boundary
condition applied therein. A very good agreement between the atomistic tight-binding and continuum model results have been obtained.
The electrostatic confinement opens perspectives for studies of localized states in the anomalous, spin and valley
quantum Hall effects conditions.

\section*{Acknowledgments}
This work was supported by the National Science Centre (NCN) according to decision DEC-2016/23/B/ST3/00821
and by the Faculty of Physics and Applied Computer Science AGH UST statutory research tasks No. 11.11.220.01/2 within the subsidy of
Ministry of Science and Higher Education. The calculations were performed
on PL-Grid Infrastructure.

 \section*{Author contributions}
B.S. performed most of the calculations. D.\.Z calculated the results for the spin-orbit coupling and the continuous version of the confinement potential. A.M.K. and B.S. analysed the 
results and wrote the paper.

 \section*{Data availability}
 All data generated or analysed during this study are included in this published article.

 \section*{Competing interests}
The authors declare no competing  interests.

\end{document}